\documentclass[12pt,a4paper,]{article}
     \let\epsilon\varepsilon
\newcommand{\comment}[1]{}

     \newcommand{\rf}[1]{(\ref{#1})}

     \newfont{\groot}{cmbx12 scaled 1200}
     \newfont{\kleiner}{cmr10 scaled 1200}

     \newcommand{\vg}[1]{\begin{equation} #1 \end{equation}}

     \usepackage{epsf,cite,alltt,}
\usepackage{graphicx,epsfig,times,mathptm,pstricks}
\usepackage{amsfonts}
\usepackage[]{amsmath}
    \usepackage{amsxtra}
    \usepackage{amstext}
    \usepackage{amssymb}
    \usepackage{latexsym}
\usepackage{helvet}

\let\epsilon\varepsilon

\usepackage{amsmath, amssymb, graphics, graphicx,amsfonts, amsthm}
 \newtheorem{theorem}{\textbf{Theorem}}
\begin{document}

\title{Natural Modes and Resonances in a dispersive stratified N-layer medium}
\author{W. Broer$^{1}$ and B.J. Hoenders$^{1}$.\\ \\
$^1$University of Groningen, Institute for Theoretical Physics\\
 and Zernike Institute for advanced materials\\
 Nijenborgh 4, NL-9747 AG Groningen, The Netherlands\\ \\
 }

\maketitle

\begin{abstract}

\noindent The properties of the natural modes in a dispersive stratified N-layer medium are investigated. Especially the focus is on the (over)completeness properties of these modes. Also the distribution of the natural frequencies
are considered. Both the degree of (over)completeness and the natural frequency distribution turn out to be
totally different from what is known for the non-dispersive case.

\end{abstract}

\section{Introduction}

\par Natural modes arise in connection with the scattering of an
incoming wave on an object. In this context, they are defined as those solutions of the scattering operator
which exist in the whole three-dimensional space $\mathbb{R}^3$,  satisfy the boundary conditions at the
surface of the finite-scatterer, and represent outgoing waves outside the medium. Natural modes were first
discovered by Cauchy in 1827 \cite{Cauchy1827},  and then later applied by Thomson \cite{Thomson},
Kol\'{a}c\v{e}k \cite{Kolacek}, and Abraham \cite{Abraham} to various scattering problems. For a
non-dispersive medium they are known to have the following properties \cite{Leung94,Leung94a,Leung97c}:

\begin{enumerate}\item They are complete within the open domain, but not
always up to the boundary. (Think of Gibbs' phenomenon of a
Fourier series, not necessarily representing the approximated
function at the endpoints.)

\item Mathematically they are `double' complete, there are two complete sets of natural modes.

\item The complex eigenvalues $k_n$ satisfy $k_n=-k_n^*$.
\end{enumerate}

\noindent In this paper, completeness of the natural modes  is understood to mean that the solution of the governing partial differential equation, subject to the boundary conditions set by the physical system, can be written as a linear combination of the natural mode eigenfunctions, i.e. this solution and the linear combination of eigenfunctions are `arbitrarily close' to each other. As these eigenfunctions satisfy the same  partial differential equation, they are a subset of $L^2_{loc}$ and hence the definition of completeness given in appendix \ref{P&Wgalore} also applies to them.
\par An example from quantum mechanics \cite{miranda} can
give a general idea of the concepts involved. From the
time-independent Schr\"{o}dinger equation

\begin{equation}(\nabla^2+k^2)\psi(\vec{r})=U(\vec{r})\psi(\vec{r})\end{equation}

\noindent and the Green's function associated with it

\begin{equation}G(\vec{r},\vec{r'};k)=\frac{e^{ik|\vec{r}-\vec{r'}|}}{|\vec{r}-\vec{r'}|}\end{equation}

\noindent we derive the corresponding scattering integral
equation:

\begin{equation}\psi(\vec{r})=\psi^{(i)}(\vec{r})-
\tfrac{1}{4\pi}\int_{V}\psi(\vec{r'})U(\vec{r'})G(\vec{r},\vec{r'};k)d^{3}r'\end{equation}

\noindent It is of fundamental importance to notice that this integral equation is \emph{not}
the standard Fredholm integral equation of the second kind because of
the \emph{non-linear} dependence of the kernel on $k$. Therefore, the
natural modes are a generalization of the results of classical
Fredholm theory: They are the solutions of the homogeneous
integral equation  in terms of which it is to be expected that the
solution of the scattering integral equation can be written inside
the domain of the scatterer.
\par Mathematically, natural mode eigenfrequencies are complex
eigenvalues of a linear or linearized differential equation,
subject to certain non-classical Sturm-Liouville boundary
conditions, see: \cite{pattanayak}, \cite{wolf73}, \cite{wolf76}, \cite{hoenders77a}, \cite{hoenders79a},
\cite{hoenders79b}. In optics (if this differential equation is the wave
equation), these eigenfrequencies correspond to the singularities
of the system, i.e. the singularities of the scattering matrix, \cite{nussenzveig}.
Physically, the natural mode formalism is a tool to describe the
\emph{energy dissipation} of a system. The imaginary parts of the
eigenvalues indicate the amount of the energy loss of the system.
(Similarly to a harmonic oscillator with damping; the imaginary
part of the frequency equals the damping coefficient). As such it
is used in various fields of physics, ranging from classical wave
mechanics, computational biophysics and mathematical physics to
general relativity and quantum gravity. (See e.g.
\cite{Beyer,Chandrasekhar&Detweiler1975} for applications in general
relativity).
\par In  the context of photonic crystals a (complex) natural mode
frequency can be related to the \emph{transmission spectrum} of the medium: the
real part of the frequency indicates the position of a resonance
peak, and its imaginary part corresponds to the full width at half
maximum of the peak. In other words: if the wavelength is
chosen to correspond to the real part of the natural mode frequency, then
the photonic crystal will transmit more electromagnetic radiation than otherwise.
\par Another important property of natural modes  in general is that
they \emph{can exist in the medium without the presence of an
incoming or driving field $\psi_{inc}$}. This can be understood in
terms of internal (electron) oscillations of the scatterer: those
oscillations will continue even after the driving field is long
gone.
\par This paper is organized as follows: the next section concerns the calculation of natural mode frequencies in the physically important special case of (non-dispersive) photonic band gap (PBG) media. The third section covers similar calculations for dispersive stratified N-layer (SNL) media: it is shown that the natural mode frequencies cluster near the resonances in the Lorentz model. The fourth section addresses the question whether the  natural mode eigenfunctions can describe the actual electromagnetic field in a dispersive (SNL) medium, (i.e. whether they are complete according to the definition given in appendix \ref{P&Wgalore}), and whether such a representation is unique (i.e. whether they are `overcomplete'). Some concluding remarks are made in the final section.

\section{Natural mode frequencies and non-dispersive N-layer media \label{naturalmodefrequencies}}

The goal of this section is to find natural mode eigenfrequencies of stratified non-dispersive n-layer media
whose respective refractive indices are assumed to be constant. These eigenfrequencies can be defined as the
singularities in the transmission and reflection coefficient of the system. (See, for instance \cite{Leung94}
and \cite{nussenzveig}) which both therefore, as we know, have the same denominators.

\par Generally, for a system of two layers or more, the natural mode frequencies cannot be found exactly.
They satisfy a transcendental equation, which can be solved numerically or for ``large" values (for more details,
see Appendix \ref{expsumroots}).
\par However, if we limit ourselves to the case of normal incidence
(so there is no angular dependence), and TE transmissions in a
periodic medium, we can find the resonance frequencies exactly in,
for instance, a system of four periods. Each period consists of
two layers each with refractive indices $n_1$ and $n_2$,
respectively. Another restriction  we would like to make is the
following: The thicknesses of the two layers $d_1$ and
$d_2$, respectively, are chosen in such a way that

\begin{equation}\label{quarterlambda}n_{1}d_{1}=n_{2}d_{2}=\tfrac{\lambda_{ref}}{4},\end{equation}

\noindent for a certain $\lambda_{ref}$. As is well known, this particular choice, which defines the class of
so-called quarter-wave stacks, simplifies to a great extent the analysis of the system at hand,
\cite{Wolter1956}. We remark in passing that these systems are optimised for reflection of pulses with centre
$\lambda_{ref}$, because the reflected waves from each layer are all exactly in phase at this wavelength. Such a medium can be used to create planar dielectric waveguides, for instance. For more details we refer to \cite{phcbook}.
\par According to Wolter
\cite{Wolter1956}, for the case of TE illumination, the numerator ($Z_m$) and the denominator ($N_m$)
of the reflection coefficient for a 2D stratified N-layer system
may be found by means of the following recursive
relations\footnote{The numerator of the \emph{transmission}
coefficient of a system with $m$ interfaces is
$2^{m}\prod\limits^{m}_{i=1}g_i$, and as noted its denominator is
identical to that of the reflection coefficient.}:

\begin{subequations}\label{Wolterrecur}
\begin{equation}Z_{m}=(g_{m}-g_{m-1})e^{-i\delta_{m-1}}N_{m-1}+(g_{m}+g_{m-1})e^{i\delta_{m-1}}Z_{m-1}
\end{equation}
\begin{equation}N_{m}=(g_{m}+g_{m-1})e^{-i\delta_{m-1}}N_{m-1}+(g_{m}-g_{m-1})e^{i\delta_{m-1}}Z_{m-1}
\end{equation}
\begin{equation}Z_1=g_1-g_0,\;\;N_1=g_1+g_0, \end{equation}
\end{subequations}

\noindent where $m$ represents the number of interfaces (see fig. 1) and the
following shorthand notations are used:

\begin{equation}g_{m}:=\tfrac{n_{m}\cos\theta_{m}}{\mu_{m}},\;\;\delta_{m}:=\tfrac{n_{m}d_{m}\omega\cos\theta_{m}}{c}.\end{equation}

\noindent These formulas apply to a general 2D stratified N-layer system.  Note that we use a different
sign convention from the one Wolter used: we assume a time dependence of $e^{-i\omega t}$, whereas he chooses
one of $e^{i\omega t}$.  Wolter's recursion formula follows from the requirement that the fields and their
derivatives must be continuous at the interfaces. Then, for a single layer ($m=2$, see Fig. 1) it follows that we have

\begin{subequations}\label{Woltersinglelayer}
\begin{equation}Z_2=(g_2-g_1)(g_1+g_0)e^{-i\delta_1}+(g_2+g_1)(g_1-g_0)e^{i\delta_1}\end{equation}
\begin{equation}\label{Woltersinglelayer1}
N_{2}=(g_2+g_1)(g_1+g_0)e^{-i\delta_1}+(g_2-g_1)(g_1-g_0)e^{i\delta_1}.\end{equation}
\end{subequations}

\noindent  The frequencies of the natural modes are the zeros of (\rf{Woltersinglelayer1}). This equation can
be solved exactly, but this is no longer possible in case of two- and more layers, (for more details,
see Appendix \ref{expsumroots}).
\par In the particular case of normal incidence we have
$\cos\theta_{m}=1$ for all integers $m$. Also there are only two
possible values of $n_m$ and only one possible value of
$n_{m}d_{m}$. As a consequence, there are only two possible values
 for $g_m$ ($\mu_{m}=1$ for all integer $m$) and only
one variable $\delta$. We will restrict ourselves to non-magnetic media, viz. $\mu_{m}=1$.
\par  As Settimi et al. \cite{Settimi2003} noted, under these (restrictive) conditions natural mode frequencies
of N-layer media can be found exactly. The natural frequencies of a 8- and 16 layer system are plotted in
the complex plane in Fig.2 and Fig.3.

\par Something similar applies to the TM case;  only the following
definition needs to be changed with respect to the TE case:

\begin{equation}\label{TM}g_{m}:=\frac{\mu_{m}\cos\theta_{m}}{n_{m}}.\end{equation}

\noindent Because we chose this medium to be non-dispersive the
refractive indices are constant, so only the values of the
parameters change, not the actual pattern of the mode frequencies.
(It is useful to remember that Settimi and Wolter use different
conventions regarding the time dependence of the oscillations,
just like in the previous case.)
\par
 As we noted before, the real part of a natural mode frequency corresponds to the
position of a resonance peak in the transmission spectrum; its
imaginary part is related to the broadness (the full width at half
maximum) of the peak. The eigenfrequency distributions shown in fig. \ref{Woltergraph}  leads to the observations
that the number of peaks  within an interval $0\leq\mbox{Re}(\delta)\leq \pi$ increases proportionally to the
number of layers (a system of $N$ layers leads to a polynomial of
degree $N$ in $e^{i\delta}$). Also, the peaks become narrower (the imaginary parts of the natural mode
frequencies are lower) in a system of sixteen layers than in a system of eight layers. However, the position
of the `gap' (i.e. interval in the spectrum without peaks) does not change. Increasing the ratio between the
two refractive indices results in lower (absolute values of)
imaginary parts of the mode frequencies, so resonance peaks in the transmission spectrum become narrower.
These results were found in \cite{Settimi2003} and confirmed by (\ref{Wolterrecur}).

\section{Calculation of natural mode frequencies in dispersive media}
\label{sectioncalcdisp}

In this section we shall investigate the consequences of allowing the medium to be temporally dispersive. The
refractive index of the $j^{th}$ layer becomes

\begin{equation}\label{dispersion}n_j^2(\omega)=1+\frac{f_j}{\omega_j^2-\omega^2-i\Gamma_{j}\omega},\end{equation}

\noindent for a characteristic frequency $\omega_j$ and a damping
coefficient $\Gamma$. The other parameter, $f_j$, is a fraction
that denotes the oscillator strength of the material, i.e. we
assume a Lorentz profile. For simplicity, the refractive indices
are assumed to have only one singularity each. The results in this section are no longer limited to quarter-wave stacks, but we still assume normal incidence.
\par We will first deal with the case of a single layer system. For
the calculation we use the scale of the characteristic frequency, in other words $\omega_j=1$. Because of
(\ref{Woltersinglelayer1}) we know that for values of the frequency close to the singularities
$\omega=\pm\sqrt{\omega_j^2-i\Gamma_{j}\omega}$, which means that $|{n_1}(\omega)| \gg 1$,  the frequencies have to
satisfy the following equation:

\begin{equation}\label{1layerdispersive}\sin(\tfrac{\omega}{c}\cdot d\sqrt{\frac{f_j}{1-\omega^2-i\Gamma_{j}\omega}})=0,
\end{equation}

\noindent  The other choices for the calculation are: $d=1$, $f_j=0.25$, $\Gamma_{j}=10^{-3}$, and
$n_0=n_2=1$. (The environment of the medium is air or vacuum). Formula \eqref{1layerdispersive} can be derived
by setting $m=2$ in Wolter's recursive formula \eqref{Wolterrecur}, and neglecting 1 with respect to
$\tfrac{f_j}{\omega_j^2-\omega^2-i\Gamma_j\omega}$ because we are near either one of the singularities of
\eqref{dispersion}.

As fig. 4 shows, the mode frequencies cluster near
the singularities of $n_1(\omega)$. Also, there are no frequencies to the
outside the outermost singularities.
\par Analogously, for a system of two layers the mode frequencies
display the same pattern: near the singularity of the refractive
index of each of the respective layers they cluster near the
aforementioned singularity. Also, in both cases, there are no mode
frequencies to either the left or the right of the outermost
singularities of the refractive indices, i.e. no eigenfrequencies
with $|\mbox{Re}(\omega)|\geq\omega_1$. Inserting $m=3$ into
Wolter's recursive formula yields the following transcendental
equation for the eigenfrequencies:

\begin{equation}\label{exacttwolayer}\begin{split}&(1+g_2)\{(g_2+g_1)(1+g_1)\exp(-i\delta_1)-
(g_2-g_1)(g_1-1)\exp(i\delta_1)\}\exp(-i\delta_2)\\&
+(1-g_2)\{(g_2-g_1)(g_1+1)\exp(-i\delta_1)+\\&(g_2+g_1)(g_1-1)\exp(i\delta_1)\}\exp(i\delta_2)=0,
\end{split}\end{equation}

\noindent where $g_{1,2}=n_{1,2}(\omega)$, $\delta_{1,2}=\tfrac{\omega}{c}
n_{1,2}(\omega)d_{1,2}$, and $g_0=1=g_3$. Now let us look for
eigenfrequencies near the resonance of the refractive index of the
first layer. This means that $|g_1|\gg 1$, $|g_1|\gg g_2$,  $g_2$
and $n_2(\omega)$ are approximately constant, and $|\delta_1|\gg 1$, hence
we obtain:

\begin{equation}-(1+n_2(\omega))\sin(\tfrac{\omega}{c} n_1(\omega) d_1)\exp(i\tfrac{\omega}{c} n_2(\omega)
d_2)+(1-n_2(\omega))\sin(\tfrac{\omega}{c} n_1(\omega) d_1)\exp(-i\tfrac{\omega}{c} n_2 (\omega)
d_2)=0.\end{equation}

\noindent Note that, in principle $n_2(\omega)$ is also frequency dependent, but this can be neglected near the resonance of $n_1(\omega)$. After some manipulations of trigonometric functions we conclude that, instead of (\ref{1layerdispersive}), the mode
frequencies  now have to satisfy

\begin{equation}\label{2layerdispersive}
\sin(n_1(\omega) d_1 \tfrac{\omega}{c})\cos(n_2(\omega) d_2 \tfrac{\omega}{c})+n_2(\omega)\cos(n_1(\omega) d_1 \tfrac{\omega}{c})\sin(n_2(\omega) d_2 \tfrac{\omega}{c})=0
\end{equation}

\noindent near the singularity of the first layer (where
$n_1(\omega)\gg1$).  The equation for the natural mode eigenfrequencies is

\begin{equation}\label{2layerdispersive2}\sin(n_2(\omega) d_2 \tfrac{\omega}{c})=0\end{equation}

\noindent near the resonance of the second layer(where $n_2(\omega)\gg1$). This formula is derived in approximately
the same way as \eqref{1layerdispersive}: the indices of refraction $n_{1,2}(\omega)$ as a function of the frequency
are given by equation (\ref{dispersion}), and neglect 1 with respect to $g_2$ in equation
\eqref{exacttwolayer}. Equation (\ref{2layerdispersive2}) is the same as (\ref{1layerdispersive}), so the
pattern of the mode frequencies must be the same as well. Near the electron resonance frequency of the first
layer $\omega_1$ we see the same clustering accordance with formula \eqref{2layerdispersive}.

\par With the aid of a famous theorem from function theory, viz. the Great Picard theorem it can be shown
that such clusterings occur always in a system with an arbitrary number of layers. The Great Picard theorem
states that an analytic function assumes every complex value, with one possible exception, infinitely
many times near an essential singularity\footnote{See appendix \ref{functiontheory}.}.
 From Wolter's recursive relation (\ref{Wolterrecur}) it can be concluded that
the functions occurring in systems like this are exponential
functions (or sines and cosines, if you prefer). In the case of TE
polarization, the dispersion model (\ref{dispersion}) gives
rise to singularities both inside and outside the arguments of the
exponential functions, so we have an equation of the type $\sum_{p}
A_{p}(\omega)\exp(iB_{p}(\omega)\tfrac{\omega}{c} d)=0$, where both $A_{p}(\omega)$ and
$B_{p}(\omega)$ are meromorphic functions of $\omega$ with singularities
in the complex $\omega$-plane. Only the singularities of the
arguments of the exponential functions (those of $B_{p}(\omega)$) are essential ones, the others
are poles. After all, $A_{p}(\omega)$ is a polynomial in the refractive indices $n_j(\omega)$ which depend on frequency according to \eqref{dispersion}, and $B_{p}(\omega)$ depends linearly on $n_j(\omega)$. According to the Great Picard theorem, clusterings like the ones we have seen in two special cases also occur more generally, in any stratified N-layer medium in the case of TE polarization.
\par If the polarization is transversal magnetic, then the
definition of the coefficients  $g_m$ changes according to
(\ref{TM}). The coefficients $\delta_m$ remain the same as in the
transversal electric case. This means that the electron resonance
frequencies are also essential singularities in the TM case.
Therefore, Picard's great theorem can be applied again and  there
will be a similar  clustering near the resonances.
\par In terms of transmission spectra, we are not quite sure what this distribution of eigenvalues means.
 Possibly, because of the positions of the natural mode eigenfrequencies in the complex
plane, the peaks may shift slightly closer to the origin  and become slightly narrower with respect to the
peak at $\pm \omega_0$ and the FWHM of $\tfrac{1}{2}\Gamma$. However, this is not the reason for calculating
such eigenfrequencies. We have done these calculations because we suspect such clusterings to represent
\emph{a complete set of modes} (See the analysis given in the next two sections).

\section{The (over)completeness of the natural modes of dispersive media}

\subsection{Introduction}

This section focuses on the question how `physical' natural modes are, i.e. whether the modes can represent
physical quantities such as wavefunctions of electromagnetic radiation. That is why we look into the
\textsl{completeness} of the natural modes. So the question is if a solution electromagnetic wave equation subject to the boundary conditions set by the stratified N-layer medium,
can be written as a linear combination of the natural mode eigenfunctions.
\par Starting with a one dimensional wave equation, Leung et al. \cite{Leung94} showed that the poles of the
Fourier transform of the Green's function, $\widehat{G}(x,y;\omega)$ correspond to the frequencies of the
eigenmodes (which Leung calls quasinormal modes). Alternatively put, the poles of  $\widehat{G}(x,y;\omega)$
are the natural mode frequencies. Also it was stated that a physically necessary and sufficient condition for
completeness of the modes is

\begin{equation}\label{completeness}\lim_{|\omega|\rightarrow\infty}\widehat{G}(x,y;\omega)=0.\end{equation}

\noindent in the lower half plane of the complex variable $\omega$.
\par However these results did not seem entirely
satisfactory, because the modes are actually \emph{overcomplete}; in other words: a representation of the wave
function or the Green's function in terms of an eigenfunction expansion is not unique. To determine `the
degree of (over)completeness'  Leung et al. \cite{Leung97c} introduced a two component formalism: A vector was
introduced with one component equal to the wave function and the other equal to the time derivative of the
wave function. From this it was concluded that the natural modes corresponding to one singularity represented
the completeness of the expansion of the wave function, and the natural modes of the singularity next to it
represented the completeness of the time derivative of the wave function.

\par The results of Leung et al \cite{Leung94,Leung94a} apply both to non-dispersive media as well as to
\textit{dispersive media.} The goal of this section is to extend the results of \cite{Leung94a} in the
following sense:  Leung et al \cite{Leung94a} showed that an overcomplete set of natural modes is generated by
the singularity at $\infty$ of the dielectric function for $\omega$, viz. $\epsilon= \epsilon_{\infty} +
\frac{\epsilon_{1}}{\omega} + \mathcal{O}{(\frac{1}{\omega^{2}})} $. We will show that \emph{each} singularity
of the dispersive dielectric function in the $\omega$ plane, which physically corresponds to a spectral line,
generates an overcomplete set of natural modes!

%In the previous section it was shown that the natural modes can be
%used to describe the Green's function of the wave equation (or
%equivalently, the wavefunction). The question still remains,
%whether all of the natural modes are needed for such a
%description, or maybe the eigenfunction expansion requires less
%modes. This section intends to address that question. To do so we
%use the same dispersion model as in section \ref{sectioncalcdisp}.
%The next sections apply the criteria for the (over)completeness of systems of functions to the natural
% modes and deal with the typical behavior of natural mode eigenfunctions, as knowledge of this behavior
% is necessary to come to statements about the (over)completeness of the natural modes.
%\subsection{Necessary and sufficient conditions  for completeness of the natural modes in 1D dispersive Stratified $N$-layer media}
\par Dispersion is traditionally phenomenologically introduced by assuming that the refractive index depends on the (time) frequency, \cite{L&L}. Therefore, we have to start in frequency space with the Helmholtz's equation :

\begin{equation}\label{Helmholtz}(\tfrac{\partial^2}{\partial x^2}+\tfrac{\omega^2}{c^2}n^{2}(x,\omega))
\widehat{\psi}(x,\omega)=0,\end{equation}

\noindent instead of with the usual wave equation (in one dimension). The `hatted' functions denote the
temporal Fourier transform of the `unhatted' functions.  Furthermore, $n^{2}(x,\omega)$ is assumed to be a
`wellbehaving' , differentiable function almost everywhere in the $\omega$-plane and to have a few
discontinuities in the $x$-direction and is supposed to be integrable with respect to $x$. Note that Eq. \eqref{Helmholtz} is the 1D Helmholtz equation, which implies we still assume normal incidence. 
\par In order to find out what (\ref{Helmholtz}) means to the system
in time space we have to apply an inverse Fourier transform:

\begin{equation}\label{memory}\tfrac{\partial^2}{\partial x^2}\psi(x,t)-\tfrac{\partial^2}{\partial t^2}
\tfrac{1}{c^2}\int\limits^{t}_{0}\rho(x,t-t')\psi(x,t')dt'=0.\end{equation}

\noindent The integro-differential equation (\ref{memory})  can be interpreted as a medium with a
memory: the whole time interval from zero to $t$ is relevant for the physics and therefore represented in the
equation of motion. For instance, in terms of the Lorentz model one can think of electrons that start to
oscillate because of the arrival of the em wave. Through these oscillations they affect the part of the wave
that has yet to enter the medium. Some books, like \cite{L&L} introduce dispersion in this way.

\subsection{(Over)completeness of the natural modes of a slab}

The problem to be addressed to in this section concerns the (over)completeness of the set of natural modes. As
we already observed before for the case of a slab that each singularity of the refractive index, viz.
$\omega=\pm\sqrt{\omega_j^2-i\Gamma_j\omega}$ leads to an infinite number of natural frequencies and natural
modes. This statement follows from the observation that close to the singularity the approximate equation to
be satisfied by the natural frequencies $\omega$, (see \rf{1layerdispersive}):

\begin{equation}\label{3layerdispersive}\sin(\tfrac{\omega}{c}\cdot
d\sqrt{\frac{f_j}{\omega_j^2-\omega^2-i\Gamma_j\omega}})=0,
\end{equation}
generates for each singularity an infinite number of roots, as the great Picard theorem tells us that near the
essential singularities $\pm\sqrt{\omega_j^2-i\Gamma\omega}$ each complex value is obtained an infinite number
of times. Hence the question arises whether these successive sets of modes are each complete or not. The key
to the solution of this problem is the analysis of the behavior of the distribution of the natural
frequencies. We refer to the books by \cite{Lewin} and \cite{P&W1967}. Before going into more detail we wish
to remark that we will freely switch between the concepts ``closure" and ``completeness", as Paley and Wiener
showed \cite{P&W1967} that these two concepts are equivalent (see appendix \ref{P&Wgalore} for definitions of
both terms).

For a slab made of dispersive material, embedded in vacuum, the natural frequencies $\omega_n$ are the roots
of \eqref{Woltersinglelayer1}, and the natural modes read as:
\vg{\sin(\tfrac{\omega_n}{c} x +\phi),}
 the natural mode eigenfunctions are therefore specific linear combinations
of functions of the form $e^{\pm i\tfrac{\omega_{n}}{c} y}$. Paley and Wiener \cite{P&W1967} studied completeness
properties of this type of functions (see appendix section \ref{P&Wgalore} for some of their most relevant
results).
\par The goal is now to apply  some of the work of Paley and Wiener in order to prove the completeness of systems of
natural modes pertaining to a single dispersive slab. More particularly, we wish to apply the theorem
\ref{P&W4}.
  \par We will now apply the Paley Wiener theorem \cite{P&W1967} for the case of a slab and construct therefore
  the canonical product $F(z)$,

\begin{equation}\label{canonicalproduct}F(z)=\prod_{m\neq
0}\Big(1-\frac{z}{\lambda_m}\Big).\end{equation}

\noindent (see \eqref{lim}). The multiplication runs over all the eigenvalues $\lambda_{m}$ of a set of natural modes,
the zeros of the canonical product must correspond to the eigenfrequencies. Then, the Paley-Wiener theorem
tells us that a set of natural modes is then, and only then, complete if the canonical product
\eqref{canonicalproduct} is \emph{not} square integrable $L^{2}(\mathbb{R})$.

\noindent Alternatively put: Furthermore, \textbf{two cases are distinguished: either the frequency is close
to one of the resonances, or its absolute value tends to infinity.} In the former case, let

\begin{equation}z:=n(\omega)\Leftrightarrow \lambda_m=n(\omega_m).\end{equation}

\noindent And if $|\omega|\rightarrow\infty$, then

\begin{equation}z:=\tfrac{\omega}{c} d\Leftrightarrow\lambda_m=\tfrac{\omega_m}{c}
d.\end{equation}

\noindent One might say that the canonical product \emph{interpolates} the eigenfrequencies, therefore $z$ is
associated with $\omega$ and $\lambda_m$ with $\omega_m$. The reason we distinguish these two cases is that both the resonances and infinity are essential singularities of this system. As we have seen in section
\ref{sectioncalcdisp} such singularities give rise to infinitely many natural modes. Therefore each of these
`clusterings' is a candidate to be a complete set of modes. (Note that, the fact that there are infinitely many
natural modes near one such singularity is not a proof that the modes are complete). \textbf{Also it follows
from the requirement that the canonical product must be in $L^2(\mathbb{R})$ if and only if it tends to zero faster than $z^{-\tfrac{1}{2}}$.} We shall take advantage of this
when investigating the  natural  modes' completeness in this section.\\

\par We will now show that that the canonical product has the following behaviour
for large values of $|z|$:

\begin{equation}\label{asymptotic}F(z)=\prod_{m=1}^{\infty}\Big(1-\frac{z}{\lambda_m}\Big)
\sim p(z)\sin(z)\end{equation}

\noindent where $p(z)$ is a polynomial in $z$ that may contain negative powers of $z$:

\begin{equation}p(z):=\sum_{j=-l}^{n}a_{j}z^{j}, \;\; l\;\mbox{and} \;n \; \mbox{are positive integers}.\end{equation}

\noindent This is the form of the canonical product in both the cases $|\omega| \Rightarrow \infty $ and $\omega
\Rightarrow \omega_{m}$, where $\omega_m$ represents the resonance frequency.

\par First let us consider the case $|\omega|\rightarrow\infty$. According to  the dispersion model \eqref{dispersion}
the refractive index can be approximated by

\begin{equation}n(\omega)=1-\tfrac{A}{2\omega^2}+O(\tfrac{1}{\omega^3}) \end{equation}

\noindent for a certain constant $A$. Wolter's formula \eqref{Wolterrecur} for one layer yields the following
equation for the eigenfrequencies

\begin{equation}(2-\tfrac{A}{2\omega^2})^2\exp(-i\tfrac{\omega}{c}(1-\tfrac{A}{2\omega^2})d)-
\tfrac{A^2}{4\omega^4}\exp(i\tfrac{\omega}{c}(1-\tfrac{A}{2\omega^2})d)=0\end{equation}

\noindent multiplying by $\tfrac{\omega^2}{A}$ gives

\begin{equation}\label{rarified}
\sin(\tfrac{1}{i}\log(\tfrac{4\omega^2}{A})+(\omega-\tfrac{A}{2\omega})\tfrac{d}{c})=0\end{equation}

\noindent which implies that the argument has to equal $m\pi$ (we
 choose the principal value for the complex logarithm). If
$|\omega|$ is chosen sufficiently large, then the term linear in $\omega$ will dominate the other terms.
Iterating once and neglecting terms of order $\tfrac{1}{\omega}$ yield

\begin{equation}\omega_m=\tfrac{c}{d}(m\pi -\tfrac{1}{i}\log(\tfrac{4m^2\pi^2}{Ad^2}))
\quad m\in\mathbb{Z}\backslash \{0\}.\end{equation}

\noindent For this case it seems appropriate to define $z=z'd$, so that $\lambda_m=\tfrac{\omega_{m}}{c}d$. 
%This scaling restricts $z'$ to the interval $0 \leq z' \leq 1 $.
From now on we will write $z$ instead of $z'$. The associated canonical product constructed from the eigenvalues for large values of $|z|$ reads as,
\cite{P&W1967}:

\begin{equation}\label{canonicalprodlargefreq}F(z)=p(z)\prod_{m\in\mathbb{Z}\backslash \{0\}}\Big(1-\frac{z}
{m\pi+i\log(\tfrac{4m^2\pi^2}{Ad^2})} \Big)
\end{equation}
It is shown in the appendix that for large values of $|z|$ the modulus of the product $L(z)$:

\begin{equation}\label{canonicalprodlargefreq2}L(z)=\prod_{m\in\mathbb{Z}\backslash \{0\}}\Big(1-\frac{z}
{m\pi+i\log(\tfrac{4m^2\pi^2}{Ad^2})} \Big)
\end{equation}
tends to a constant. Hence combination of \rf{canonicalprodlargefreq}  and \rf{canonicalprodlargefreq2} leads to:

\begin{equation} \label{canonicalproductestimate} |F(z)|\sim |p(z)|\cdot\mbox{const.}\neq 0.\end{equation}

\noindent Depending on the polynomial $p(z)$ each resonance generates either  a (over)complete set of modes if the polynomial contains only positive powers of $z$, or a set of modes which is not complete if the polynomial contains only negative powers of $z$. However, if this is the case, such an incomplete set but can be made complete by the addition of only a \emph{finite} number of modes generated by the one of the other singular points. \par This means that for a 1D dispersive stratified $N$-layer
medium, the natural modes are \emph{at least} $N+1$-fold complete, if it is assumed that each layer has two
resonances. Alternatively put,  the Green's function or the wave  function  for a photonic crystal with $2N$
resonances can be expanded in terms of natural mode eigenfunctions in at least $N+1$ ways.
\par This completes the proof of \eqref{asymptotic}.

\par Writing $F(z)$ like  equation \eqref{asymptotic} simply means
that we include eigenfrequencies that are still in the vicinity of one of the singularities (either at one of
the resonances or at infinity), but not close enough to one of them to display the pattern $\lambda_m=m\pi$.
 \par Equation \eqref{estimate} can be verified as follows:
 Let $\epsilon\rightarrow\infty$, $n=0$ and $A=\epsilon$ then
 because of the form \eqref{asymptotic} we can write
 $F(y+i\epsilon)\sim \exp(\epsilon)$. Also it is clear that condition \eqref{lim} holds:
 $\lambda_m$ is at most linear in $m$. So theorem \ref{P&W4} can
 be applied.
\par As  the eigenfrequencies $\omega_m$ cannot be determined
exactly, the precise form of $p(z)$ also remains unknown. The modes that display the pattern $\lambda_m=m\pi$
are \emph{not} complete: they yield $F(z)=\tfrac{\sin z}{z}$ which is in $L^2(\mathbb{R})$. However, these
modes correspond to the case of a clamped string without the important eigenmode $z=0$ (no oscillation). They
do not aptly describe this system as it is open, unlike a clamped string. This does tell us that the modes are
not complete if we take the neighborhood around one of the resonances too small. Depending on the asymptotic
behavior of $p(z)$, we distinguish two possibilities

\begin{enumerate}
\item\label{case1} $p(z)$ only contains negative powers of $z$. As
$F(z)$ is an entire function, a sufficient condition for $F(z)$ to be in $L^{2}(\mathbb{R})$ is

\begin{equation}\lim_{x\rightarrow\infty}\sqrt{x}F(x)=0\end{equation}

\noindent where $x$ is real. So in this case the canonical product is in $L^2(\mathbb{R})$, and according to
theorem \ref{P&W4} the natural modes are not complete. The same theorem also states that in this case, the
corresponding set of eigenfunctions can be made complete by the adjunction of a finite number of functions of
a similar form. To this end, some other natural mode eigenfunctions can be used. However, there is no physical
reason to prefer one natural mode to another. Neither is there a physical reason why, say, $q$ natural modes
are required to make them complete and not another number.

\item\label{case2} $p(z)$ contains at least one positive power of
$z$. Then $F(z)\notin L^{2}(\mathbb{R})$ and the natural modes in this area are complete.
\end{enumerate}

Whichever possibility is the correct one, from physical considerations it seems that the natural modes in the
neighbourhood of the resonances are complete anyway. If more layers are added to the system, more resonances will occur. This means there will also be more clusterings of natural modes near these resonances. So, a
system of $N$ slabs is at least $N$-fold complete, even if $p(z)$
contains only negative powers of $z$.\\

\section{Conclusions}

The analysis of the pertinent properties of these fundamental modes,
\emph{to be considered as the most ``natural'' set of functions for the expansion of the field,} is
 of paramount interest. The completeness property of the field is especially one of the most
important and interesting features of these modes to be studied. In some special cases, for instance in a non-dispersive, periodic 1D SNL medium with quarter-wave stacks, the natural modes formalism is an efficient tool to reveal information about
transmission spectra of such media. 
\par In \cite{Leung94}, Leung et al. showed that, for a system described by the  1D wave
equation without dispersion, a sufficient (and possibly necessary) condition for completeness of the natural
modes is that the Fourier transformed Green's function vanishes for sufficiently large frequencies. The
generalization to dispersive media is relatively straightforward: the same condition applies, although
the expression for eigenmode expansion coefficients is slightly more complicated \cite{Leung94a}. This result is obtained
without outlining a specific dispersion model.
\par In order to investigate the degree of (over)completeness of
natural modes in 1D photonic crystal we have chosen the following dispersion model:

\begin{displaymath}n_j(\omega)=\sqrt{1+\frac{f_j}{\omega_j^2-\omega^2-i\Gamma_{j}\omega}}\end{displaymath}

\noindent for the $j$th layer of the medium. If each layer of the photonic crystal is assumed to have two
resonance  frequencies, then the natural modes in a medium of $N$ layers is at least $N+1$ fold complete. As
of yet, we are unsure of what this tells us about either the natural mode formalism or the used dispersion
model.
\par (Over)completeness of the natural mode expansion may imply that natural modes are `physical' in a certain way, but it does not mean that they are useful (we may still need an unpractically large number of modes to aptly describe our system). Based on
\cite{Hoenders2005} we suspect that, in the typical photonic crystal region, where the wavelength of
the electromagnetic radiation is of the same order of magnitude as the thickness of a layer, that
electromagnetic wave couples to only a few modes. Currently, however, there is still no proof of this.
\par Our results were derived for dispersive SNL media with normal incidence. To generalize to 2D systems (i.e. for $\theta_m\neq 0$ in fig. 1) Wolter's recursive formula \eqref{Wolterrecur} can still be used. Obviously the angle of incidence $\theta_m$ does not depend on frequency but the angles of refraction do, which would complicate such an analysis. However it seems likely that the Lorentz resonances also form essential singularities in this case, and hence the natural mode frequencies also cluster near the resonances. (The only way for this \emph{not} to happen would be if the frequency dependence of the angles of refraction somehow removed the essential singularities). So in 2D we would expect the same degree of `overcompleteness' to occur as in 1D. In 3D, Wolter's formula is no longer valid because both the TM and the TE modes contribute to the electromagnetic pulse. In this case, the transfer matrix method \cite{phcbook} can be used to investigate the completeness properties of the natural modes.
\par We gratefully acknowledge useful discussions with R. Uitham, M. Bertolotti, and A. Settimi.

\appendix

\section{Mathematical Theorems on the properties of functions}\label{math}

\subsection{Roots of exponential sums}\label{expsumroots}

Equation (\ref{Wolterrecur}) shows  that trying to find the
natural mode frequencies leads to a transcendental equation ( more
specifically, an exponential sum). Mathematicians studied roots of
such equations in the 1930s. In this appendix, some useful
theorems and results will be given (without proof). For more
details see \cite{Langer29}, \cite{Langer31}, and
\cite{MacColl1934}.\\
\par Langer \cite{Langer29} derived the following theorem:

\begin{theorem}\label{Langer1}If the constants $B_j$ are real and

\begin{equation}0=B_{0}<B_1<...<B_J\end{equation}

\noindent then for $|\rho|$ sufficiently large the roots of the
equation

\begin{equation}\label{expsum}\sum\limits^{J}_{j=0}[b_j]e^{\rho
B_j}=0, b_0\neq0, b_J\neq0\end{equation}

\noindent lie in the strip bounded by the lines

\begin{equation}\mbox{Re}(\rho)=\pm c,\end{equation}

\noindent where $c$ is a suitably chosen real constant. The number
$N$ of roots lying in any interval of this strip of length $l$
satisfies the relation

\begin{equation}B_{J}l/(2\pi)-(J+1)\leq N \leq B_{J}l/(2\pi)
+(J+1).\end{equation}

\noindent Moreover if $\rho$ remains uniformly away from the zeros
of (\ref{expsum}) the left hand member of the equation is
uniformly bounded from zero.
\end{theorem}

\noindent This result is useful because it concerns exponential sums of the type we  have encountered in
section \ref{naturalmodefrequencies} and\ref{sectioncalcdisp}. Because the imaginary unit $i$ occurs in
arguments of the exponents of the sum in (\ref{Wolterrecur}), the imaginary parts of the natural mode
frequencies are bounded. Also Theorem \ref{Langer1} tells us that the number of roots is proportional  to the
length of the strip. This implies that there are infinitely many natural modes in the entire complex plane (
if we stay sufficiently far away from the origin).\\

\par The next result, which Langer \cite{Langer31} obtained a few years later, is also relevant
 to us:

\begin{theorem}If in the exponential sum

\begin{equation}\label{expsum2}\Phi(z)=\sum\limits_{j=0}^{n}A_{j}e^{c_{j}z}\end{equation}

\noindent the coefficients are constant and the exponents
commensurable (the arguments of the exponents are integer
multiples of each other), the sum becomes of the form

\begin{equation}\Phi(z)=\sum\limits_{j=0}^{n}a_{j}(e^{\alpha
z})^{p_{j}}, p_0=0,p_j\in \mathbb{N}\end{equation}

\noindent and the distribution of the zeros is given explicitly by
the formula

\begin{equation}z=\tfrac{1}{\alpha}(2m\pi i+\log
\xi_j),\xi_{j}:= e^{\alpha z}\end{equation}

\noindent where $m \in \mathbb{Z}$ and $j$ is a natural number
$\leq p_n$.
\end{theorem}

\noindent This may seem complicated at first but the special case of commensurable exponents is actually
theoretically the simplest one, because it makes the problem of the distribution of the zeros essentially an
algebraic one. This is also the theorem we used for the calculations in section \ref{naturalmodefrequencies}
and\ref{sectioncalcdisp}. The assumption of commensurability is no doubt a limiting one, but it includes a few
important special cases, such as a trigonometric sum (a partial sum of a Fourier series).

\par Also, without assuming commensurability, Langer \cite{Langer31} showed that theorem
\ref{Langer1} is not only valid far away from the origin, but
everywhere in the complex plane.

\subsection{Some theorems from function theory}\label{functiontheory}

\begin{theorem}\textbf{(`Great Picard Theorem.')}\label{GreatPicard}
Suppose an analytic function $f(z)$ has an essential singularity
at $z=a$. Then in each neighborhood of $a$, $f(z)$ assumes each
complex value, with one possible exception, an infinite number of
times.\end{theorem}

Proof and corollaries can be found in most textbooks on function
theory, like \cite{Conway1978}.
\par This is the type of singularity we encountered in section
\ref{sectioncalcdisp}. The analytic function in this case is the denominator of the reflection coefficient.
The essential singularity is the (electron) resonance frequency. Of course zero includes the complex values
this function assumes infinitely many times near the essential singularity, which explains the clustering of
roots displayed in figure \ref{cluster1}.
\par The `one possible exception' is any function of the form
$e^{1/z}$ near $z=0$. This type of function cannot assume zero
since it has no roots.
\par We also require the

\begin{theorem}\textbf{(`Weierstrass factorization
theorem.')}\label{Weierstrassfact} Let $f$ be an entire function
and let $\{a_n\}$ be the non-zero zeros of $f$ repeated according
to multiplicity; suppose $f$ has a zero of order $m\geq 0$ (a zero
of order $0$ at $z=0$ means $f(0)\neq 0$). Then there is an entire
function $g$ and a sequence of integers $\{p_n\}$ such that

\begin{equation}f(z)=z^{m}e^{g(z)}\prod\limits_{n=1}^{\infty}E_{p_n}\Big(\frac{z}{a_n}\Big),
\end{equation}

\noindent where for every natural number $p$

\begin{equation} \begin{split}& E_{0}(z):=1-z
\\& E_{p}(z):=(1-z)\exp(z+\tfrac{z^2}{2}+...+\tfrac{z^p}{p}),\quad
p\geq 0\end{split}
\end{equation}

\noindent The numbers $p_n$ are chosen in such a way that the series:

\vg{\sum_{n}\left( \frac{z}{|a_{n}|}  \right)^{1+p_{n}}  <\infty  }

\end{theorem}

The function $E_p(z)$ is known as the \emph{elementary factor}.
Note that $E_{p}(\tfrac{z}{a})$ has a simple root at $z=a$ no
other roots. In a way this is a generalization of Gauss' main
theorem of algebra about the factorization of polynomials.

\subsection{Equivalence of closure and completeness in
$L^2_{loc}$}\label{P&Wgalore}

The set of functions $\{f_n(x)\}\subset L^{2}(a,b)$
($a,b\in\mathbb{R},\quad a>b$) is said to be \textsl{closed over
$(a,b)$} if

\begin{equation}\int\limits_{a}^{b}f(x)f_n(x)=0
\end{equation}

\noindent implies $f(x)\equiv 0$  almost everywhere on
$(a,b)\quad\forall f(x) \in L^{2}(a,b)$. The set of functions
$\{f_n(x)\}$ is said to be \textsl{complete} if $\forall f(x)\in
L^{2}(a,b), \varepsilon >0$ there is a polynomial

\begin{equation}P_n(x)=\sum\limits_{1}^{n}a_{k}f_{k}(x)\end{equation}

\noindent such that

\begin{equation}\int\limits_{a}^{b}|P_n(x)-f(x)|^2<\varepsilon\end{equation}

\noindent For all practical intents and purposes, we can think of this as $P_n(x)=f(x)$, because in our
application $f(x)$ is a solution of a differential equation, so this more general formulation is not needed
for it. In \cite{P&W1967} Paley and Wiener proved that

\begin{theorem}\label{P&W1}
A set of functions $\{f_n(x)\}\subset L^{2}(a,b)$  is closed over
$(a,b)$ if and only if it is complete.
\end{theorem}

This theorem isn't hard to intuitively picture: It states that the
only square integrable function that is orthogonal to all
functions of a complete set is the function that is identical to
zero. Analogously we can imagine that the only vector
perpendicular to all the vectors of a complete set of vectors is
the vector with length zero.
\par We will need this relationship between closure and
completeness  for the theorems \ref{P&W4}. Also it will be assumed
that

\begin{equation}\label{lim'}\lim_{n\rightarrow\infty}\tfrac{\lambda_n}{n}=1.\end{equation}

\noindent In this case the entire function

\begin{equation}F(z)=\prod\limits_{n=1}^{\infty}(1-\tfrac{z^2}{\pi^2
\lambda^{2}_{n}})\end{equation}

\noindent exists according to theorem \ref{Weierstrassfact}. (This statement follows from the conditions for
this theorem \ref{Weierstrassfact}, taking $p_{n}=0$).

\begin{theorem}\label{P&W2}Let \eqref{lim'} be true.
Furthermore, let $F(z)\in L^{2}(\mathbb{R})$. Then the set of
functions $\{e^{\pm i\lambda_{n}x}\}$ cannot be closed over
$(a,b)$. Again, let $zF(z)\in L^{2}(\mathbb{R})$. Then the set of
functions $\{1,e^{\pm i\lambda_{n}x}\}$ cannot be closed on
$L^{2}(a,b)$. In either case, a finite number of the functions of
the set may be replaced by an equal number of other functions of
the form $e^{i\lambda x}$.
\end{theorem}

The for this application main theorem on the completeness (closure) of sets of functions of the exponential type \cite{P&W1967} is
given below:

\begin{theorem}\label{P&W4}Let

\begin{equation}\label{lim}\lim_{m\rightarrow\infty}\frac{|\lambda_m|}{m}=1\end{equation}

\noindent then according to the Weierstrass factorization theorem (see theorem \ref{Weierstrassfact} in section
\ref{functiontheory} from appendix \ref{math}) the following entire function exists:

\begin{equation}F(z)=\prod^{\infty}_{m=1}(1-\tfrac{z^2}{\pi^2 \lambda_{m}^{2}})\end{equation}

\noindent and let

\begin{equation}\label{estimate}
|F(y+i\epsilon)|\geq\frac{A}{1+|y|^n}>0\end{equation}

\noindent for all real $y$, some $A>0$ , and some $\epsilon$ and
$n$. Then the set of functions $\{e^{\pm i\lambda_m y}\}$ will be
closed or not closed on $L^{2}(a,b)$ according as $F(z)$ does not
or does belong to to $L^{2}(\mathbb{R})$. It can always be made
closed by the adjunction of a finite number of of functions
$e^{i\lambda y}$. The set of functions $\{1,e^{\pm
i\lambda_{m}y}\}$ will be closed or not closed on
$L^{2}(\mathbb{R})$ according as $zF(z)$ does or does not belong
to $L^{2}(\mathbb{R})$.
\end{theorem}

\par Another relevant result by Paley and Wiener is the
 following:

 \begin{theorem}\label{P&W3} Let \eqref{lim'} be true and let the set of
functions $\{e^{\pm i \lambda_{n}x}\}$ be closed on $L^{2}(a,b)$
but let it cease to be closed on the removal of some one term.
Then it ceases to be closed on the removal of any one term,
$F(z)\not\in L^{2}(\mathbb{R})$, but $F(z)\in L^{2}(1,\infty)$.
Again, if the set of functions $\{1,e^{\pm i \lambda_{n}x}\}$ is
closed on $L^{2}(a,b)$, but ceases to be closed on the removal of
some one term, this term is arbitrary, then $zF(z)\not\in
L^{2}(\mathbb{R})$ but $F(z) \in L^{2}(\mathbb{R})$.
\end{theorem}

\section{The derivation of \rf{canonicalproductestimate} }

We start from the product $L(z)$:

\begin{equation}\label{canonicalprodlargefreq1}L(z)=\prod_{m\in\mathbb{Z}\backslash \{0\}}\Big(1-\frac{z}
{m\pi+i\log(\tfrac{4m^2\pi^2}{Ad^2})} \Big)
\end{equation}

\noindent Taking the logarithm of both the left hand side and the
right hand side of equation \eqref{canonicalprodlargefreq} yields

\begin{equation}\label{Riemannlimit} \log(L(z))\sim P\int\limits_{-\infty}^{\infty}\log\Big(1-\frac{z}{m\pi+
i\log(\tfrac{4m^2\pi^2}{Ad^2})} \Big)dm\end{equation}

\noindent where $P$ denotes the Cauchy principal value. The
`$\sim$' changes into an equal sign if the steps in the canonical
product are small enough. Then $\log(L(z))$ becomes a Riemann sum
of the integral occurring on the right hand side of
\eqref{Riemannlimit}.
\par Integration by parts changes the integrand into a fraction:

\begin{equation}P\int\limits_{-\infty}^{\infty}\frac{mz(\pi+i\tfrac{2\pi^{2}}{Ad^{2}m})dm}
{[m\pi+i\log(\tfrac{m^{2}\pi^{2}}{Ad^{2}})]^{2}+z(m\pi+i\log(\tfrac{m^{2}\pi^{2}}{Ad^{2}}))}
\end{equation}

\noindent Substituting
$p_{m}:=m\pi+i\log(\tfrac{m^{2}\pi^{2}}{Ad^{2}})$ simplifies this
integral considerably:

\begin{equation}\log(L(z))\sim P\int\limits_{-\infty}^{\infty}\frac{m\tfrac{d p_{m}}{dm}zdm}
{p_{m}(p_{m}+z)}
\end{equation}

\noindent The integrand has two poles that correspond to
$p_{m}=-z$ :

\begin{equation}
m'_{\pm}(z):=\tfrac{2i}{\pi}W(\pm\tfrac{i}{2}d\sqrt{A}\exp(\tfrac{iz}{2})),
\end{equation}

\noindent and two poles at $p_m=0$:

\begin{equation}
m'_{\pm}(z=0)=W(\pm\tfrac{i}{2}d\sqrt{A})
\end{equation}

\noindent where $W$ denotes the (principal value of the)
\emph{Lambert W-function}\cite{LambertW,LambertWphys}. This
function is defined as the  multi-valued solution $W(z)$ of the
equation $z=W(z)\exp(W(z))$. The principal value solution is one
of the two real branches. As we chose the principal value of the
complex logarithm, we also find the poles in terms of the
principal value of the Lambert W-function. The residues near these
poles are

\begin{equation}-\tfrac{2i}{\pi}(m_{+}(z)+m_{-}(z)+m_{+}(z=0)+m_{-}(z=0))
-\tfrac{4i\pi^{2}}{Ad^{2}}.\end{equation}

\noindent The terms with $m_{\pm}(z=0)$ correspond to the
contribution due to the two poles at $p_{m}=0$. Multiplying this
by $2\pi i$ yields the outcome of the integral. Because we have
chosen the principal value of $\log(L(z))$ we wish to estimate the
modulus of this outcome. In order to do so we use the relation
$|W(z)|=W(|z|)$:

\begin{equation}0<|\tfrac{8}{\pi}(m_{+}(z)+m_{+}(z=0))+\tfrac{8\pi^3}{Ad^2}|\leq
\tfrac{16}{\pi}W(\tfrac{1}{2}\sqrt{A}d)+\tfrac{8\pi^3}{Ad^2}\end{equation}

\noindent which does not depend on $z$! So the absolute value of
the product \eqref{canonicalprodlargefreq1} behaves as:

\begin{equation}\label{absvaluel(z)}|L(z)|\sim\mbox{const.}\neq 0.\end{equation}

%\bibliography{c:/publicatie/bib/litabbr,c:/publicatie/bib/books,c:/publicatie/bib/maarten,c:/publicatie/bib/photonic,c:/publicatie/bib/aanvr,c:/publicatie/bib/transport,c:/publicatie/bib/rheingold}
%\bibliographystyle{c:/publicatie/bib/myamsplain}

 \newpage
 Caption Fig. 1.:\vspace{2cm}
 
 Sketch of the parameters and geometry of the problem. There are $m$ interfaces and $m-1$ slabs. As we assume normal incidence ($\theta_m=0$) there is no angular dependence.\vspace{2cm}
 
 Caption Fig.2.: \vspace{2cm}

Natural mode frequencies of two non-dispersive,
1D SNL media quarter-wave stacks. The polygons represent the mode frequencies of an eight layer system
and the asterisks represent those of a sixteen layer system. the
ratio between the refractive indices of the two layers is
$\tfrac{n_1}{n_0}=1.5$ in both cases. \vspace{2cm}

 Caption Fig.3.: \vspace{2cm}

Natural mode frequencies of a  non-dispersive, 1D SNL medium with four  quarter-wave stacks. The ratio between the
refractive indices is in this case
$\tfrac{n_1}{n_0}=2$ \vspace{2cm}

 Caption Fig.4.: \vspace{2cm}

The mode frequencies of a single layer system cluster near
the singularities of the refractive index at $(\pm 1,-0.0005)$.
 The mode frequencies are symmetrical around the imaginary axis. This graph is in units $\omega_0$
 
\newpage

\begin{figure}[h]\label{m_interfaces}
\includegraphics[width=14cm]{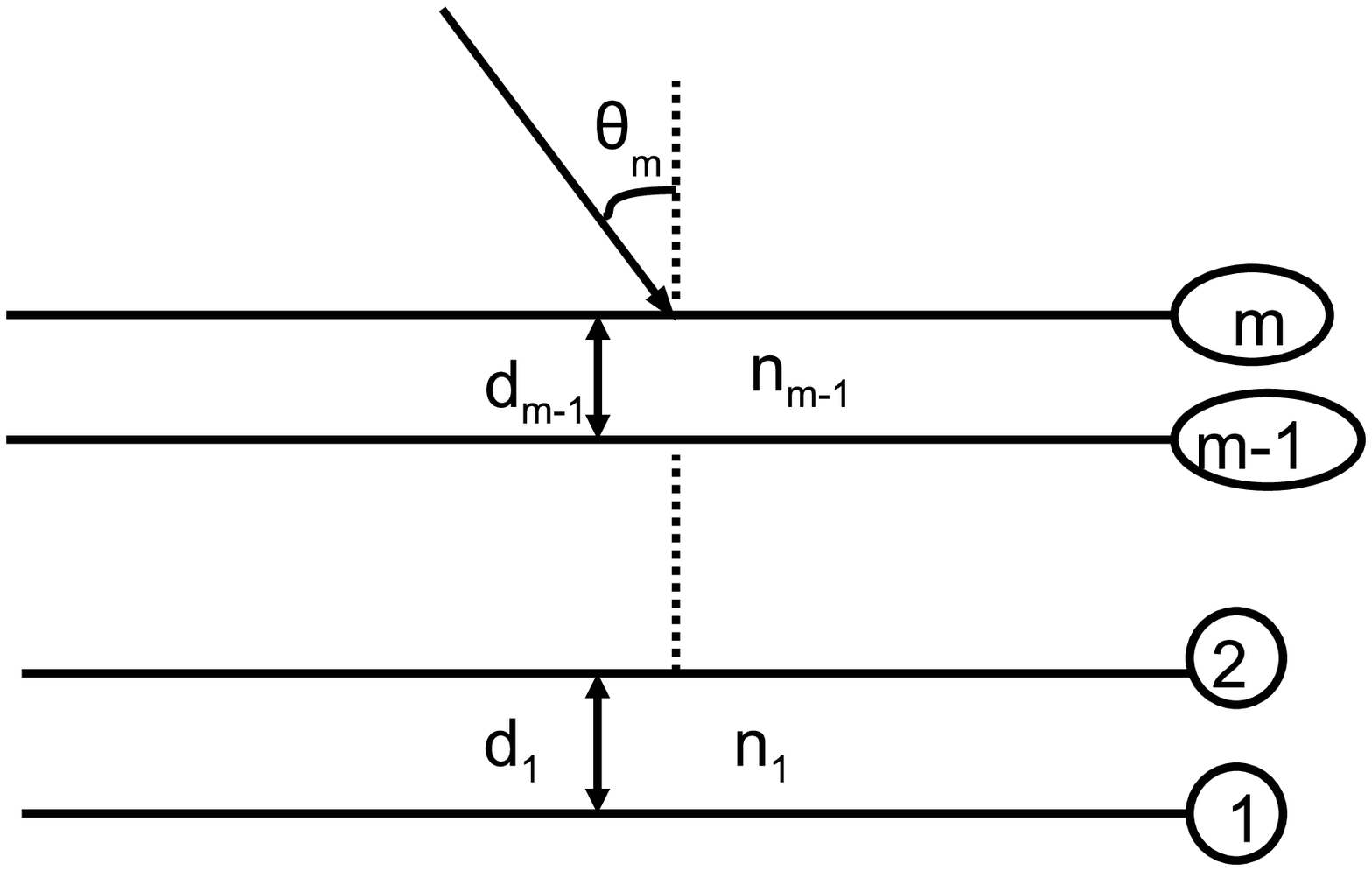}\end{figure}

\newpage

\begin{figure}[h]\label{Woltergraph}
\includegraphics[width=14cm]{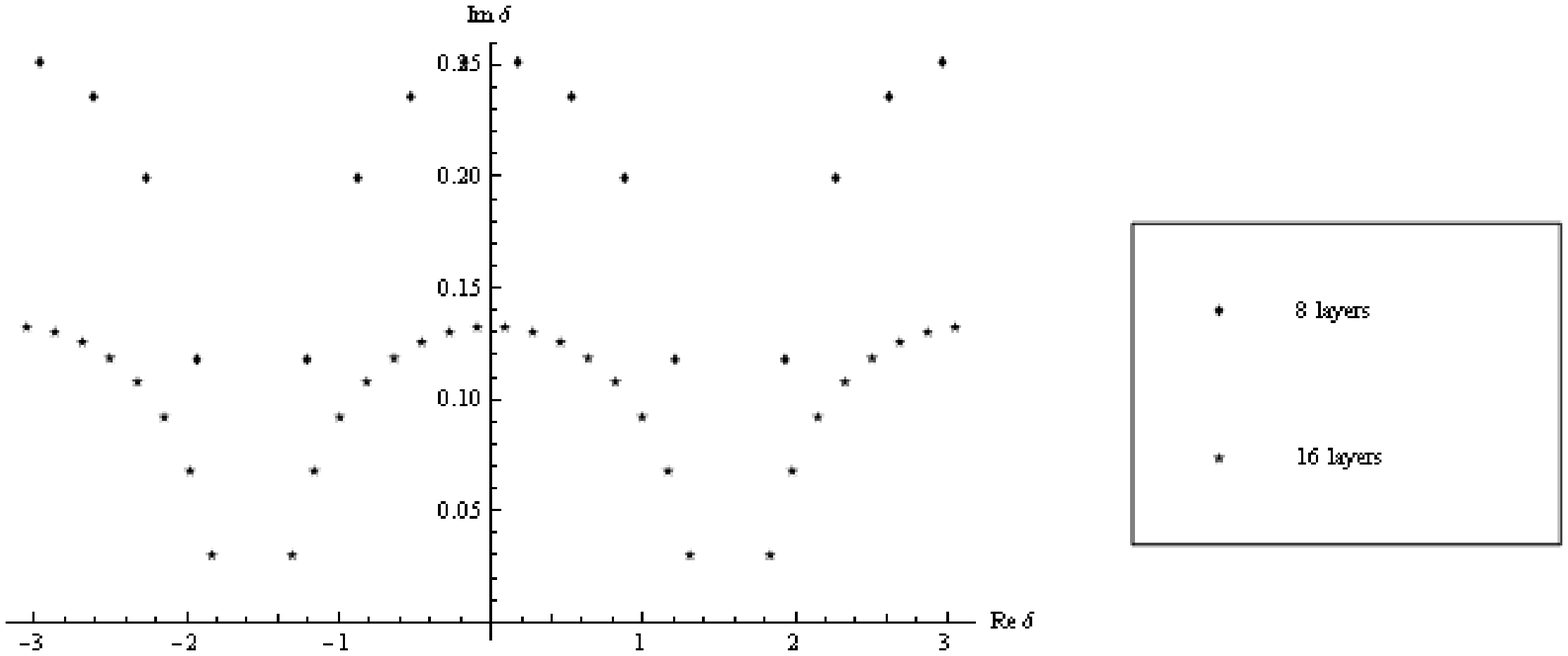}\end{figure}

\newpage

\begin{figure}[h]
\includegraphics[width=14cm]{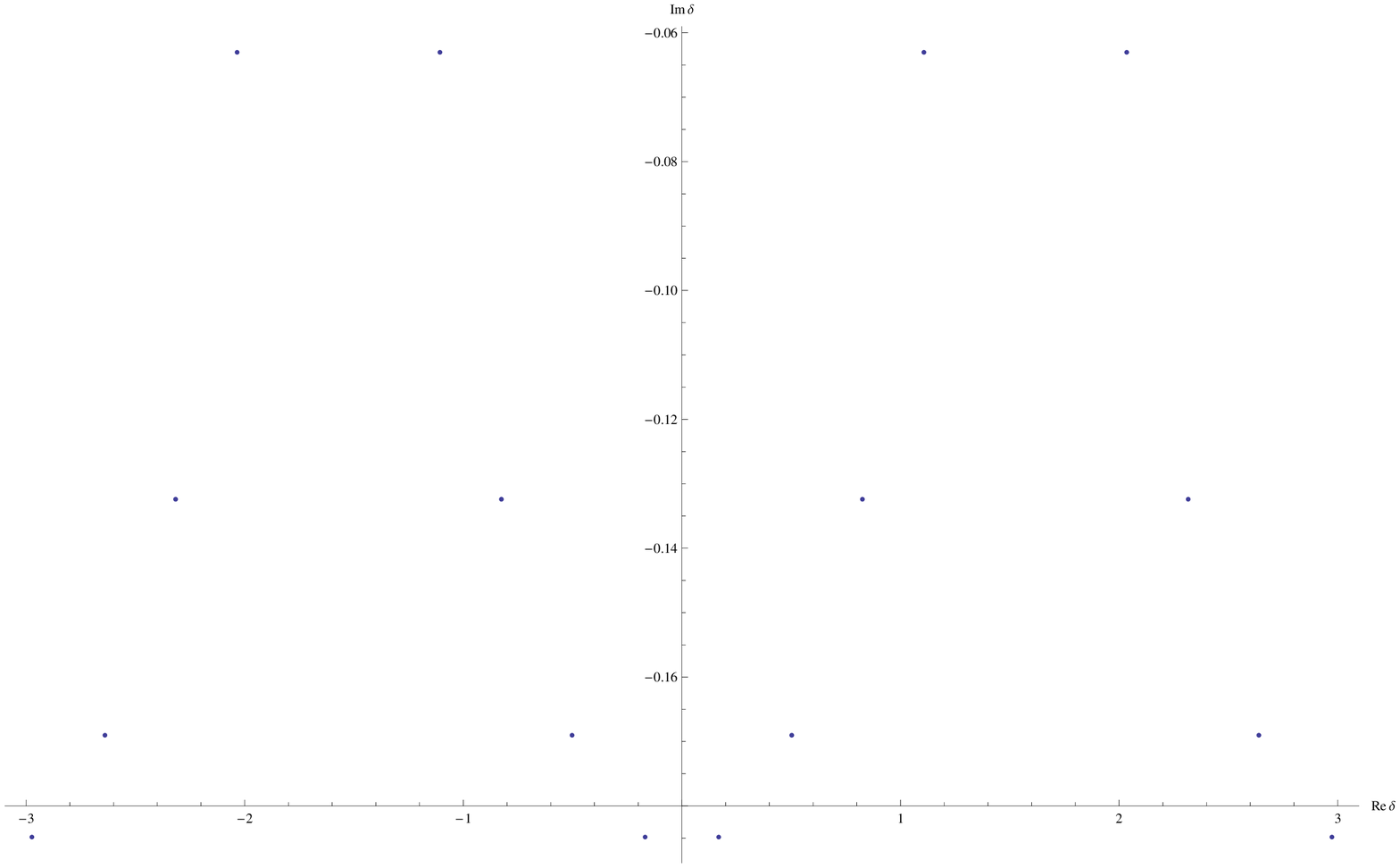} \label{bleh} \end{figure}

\newpage

\begin{figure}[h]\label{cluster1}\includegraphics[width=14cm]{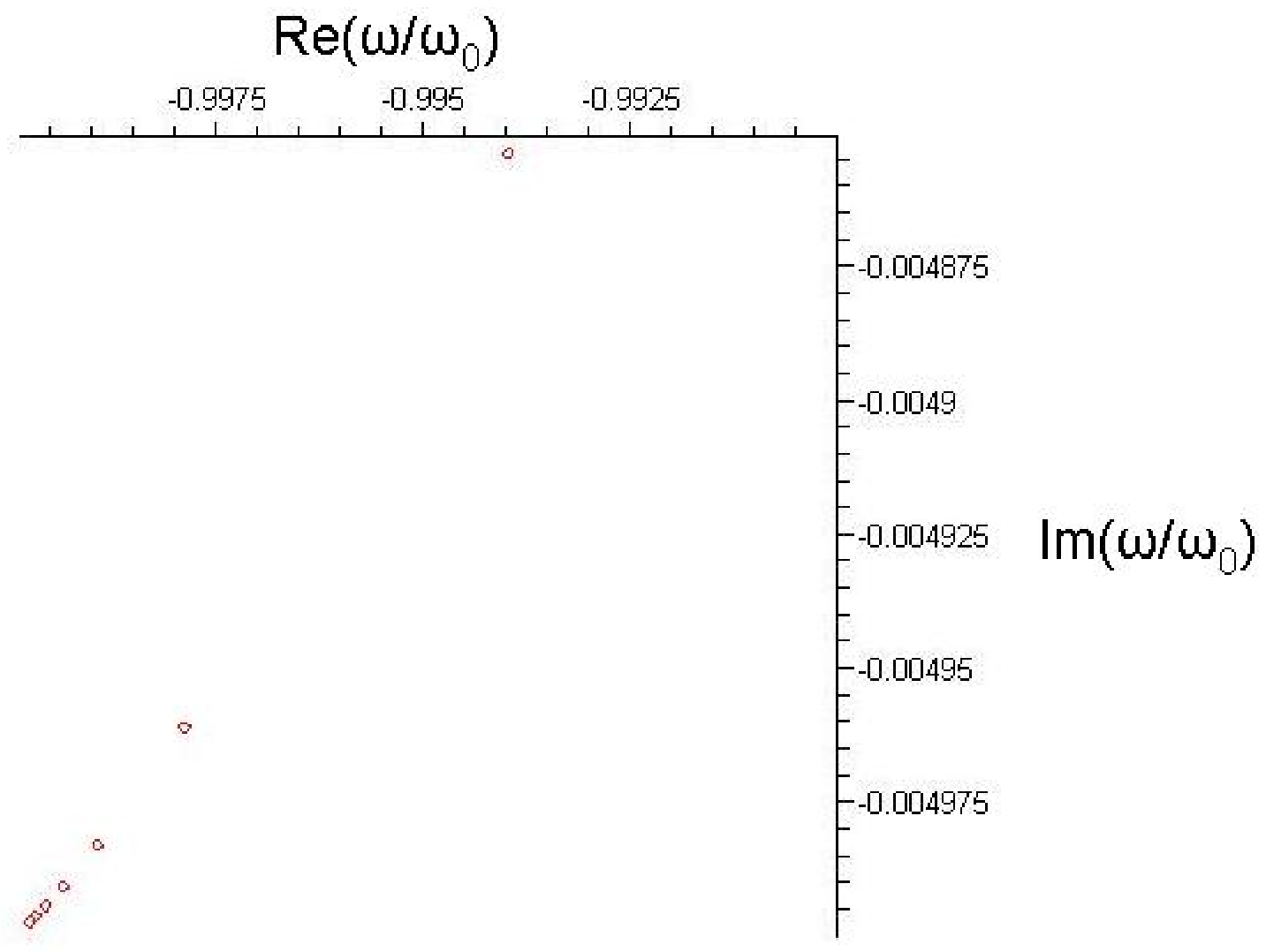}
\end{figure}

\end{document}